\documentclass[12pt,preprint]{aastex}

\slugcomment{}

\shorttitle{From Prestellar Cores to Protostellar Cores}
\shortauthors{Aikawa et al.}

\begin{document}


\title{Molecular Evolution and Star Formation:\\
From Prestellar Cores to Protostellar Cores}


\author{Yuri Aikawa}
\affil{Department of Earth and Planetary Sciences, Kobe University,
Kobe 657-8501, Japan}

\author{Valentine Wakelam}
\affil{Universit\'e Bordeaux I, LAB, UMR5804, 
BP 89, 33270 Floirac, France }

\author{Robin T. Garrod}
\affil{Max-Planck-Institut f\"{u}r Radioastronomie,
Auf dem H\"{u}gel 69, 53121 Bonn, Germany}

\and

\author{Eric Herbst}
\affil{Departments of Physics, Chemistry, and Astronomy, The Ohio
State University, Columbus, OH 43210}



\begin{abstract}
We investigate molecular evolution in a star-forming core that is initially
a hydrostatic starless core and collapses to form a low-mass protostar.
The results of a one-dimensional radiation-hydrodynamics calculation are adopted
as a physical model of the core. We first derive radii at which CO and large
organic species sublimate. CO sublimation in the central region starts shortly
before the formation of the first hydrostatic core.
When the protostar is born, the CO sublimation radius extends to 100 AU, and the
region inside $\lesssim 10$ AU is hotter than 100 K, at which some large
organic species evaporate.
We calculate the temporal variation of physical parameters in infalling shells,
in which the molecular evolution is solved using an updated gas-grain chemical
model to derive the spatial distribution of molecules in a protostellar core.
The shells pass through the warm region of $10 -100$ K in several $\times$
$10^4$ yr,
and fall into the central star $\sim 100$ yr after they enter the region
where $T \gtrsim 100$ K. We find that large organic
species are formed mainly via grain-surface reactions at temperatures of $20
-40$ K and then desorbed into the gas-phase at their sublimation temperatures.
Carbon-chain species can be formed
by a combination of gas-phase reactions and grain-surface reactions
following the sublimation of CH$_4$.
Our model also predicts that CO$_2$ is more abundant in isolated cores,
while gas-phase large organic species are more abundant in cores embedded in
ambient clouds.
\end{abstract}

\keywords{stars: formation --- ISM: molecules --- ISM: clouds --- ISM:
individual (IRAS 16293-2422, L1527)}

\section{Introduction}

In a last decade, great progress has been made in our understanding of the
chemical evolution of low-mass star-forming cores \citep[][and references
therein]{fra07, bt06}.  Among the observational advances are the detection of 
chemical fractionation in several prestellar cores; emission lines of the rare
isotopes of CO are weaker at the core center than at outer
radii, while emission lines of nitrogen-bearing species (e.g., N$_2$H$^+$)
show relatively good correlation with the centrally-peaked dust continuum
\citep[e.g.][]{cas99, taf02}.

Theoretical models show that the CO depletion is caused by adsorption onto
grains \citep{bl97,aik01}, and that the CO depletion helps to temporarily
maintain the N$_2$H$^+$ abundance at the core center. Theoretical
models also predict that a fraction of the adsorbed CO will be hydrogenated
to form H$_2$CO and CH$_3$OH on grain surfaces \citep{ar77,hhl92}, which is
confirmed by laboratory experiments \citep{wkl02,fuchs07}. The existence of
solid methanol in low-mass star formation regions has been confirmed
observationally; \citet{ppp03}
detected a high abundance of CH$_3$OH ice (15-25 \% relative to water ice)
towards three low-mass protostars  among $\sim 40$ observed protostars. Radio observations find gaseous CH$_3$OH
to be abundant in the central regions of protostars \citep{sjv02}.
Since the formation
of CH$_3$OH is inefficient in the gas phase \citep{gpc06,gwh06}, it must be formed
by the hydrogenation of CO on grain surfaces in the prestellar core stage,
and then sublimated to the gas-phase as the core is heated by the protostar.
Although CH$_3$OH ice is not detected towards the majority of low-mass protostars
\citep{ppp03} and the background star Elias 16 \citep{chi96}, the upper limit for the
CH$_3$OH ice abundance is a few $\%$ relative to water ice, which is not low enough
to contradict the idea that gaseous CH$_3$OH around protostars is originally
formed by grain-surface reactions and then sublimated.

Two major questions, however, remain to be answered, the first being at what stage
CO returns to the gas phase. The
core remains nearly isothermal as long as  cooling by radiation is more
efficient than heating by contraction (compression). Eventually, though,
the heating overwhelms the cooling, so that the core center becomes warmer.
A newly-born protostar further heats the surrounding core. 
Laboratory experiments show although some fraction of CO can be entrapped in water
ice \citep{cdf03}, a significant amount of CO sublimates at around 20 K
\citep{sa88}. In order to predict if an observable amount of CO returns
 to the gas-phase during the prestellar core stage or after the birth of
a protostar, the temperature distribution in a core should be calculated by
detailed energy transfer. Such a prediction is important in order to 
observe very young protostars.  Once it is sublimated, CO again becomes a
useful observational probe. In addition, CO sublimation significantly affects
the gas-phase chemistry; for example, it destroys N$_2$H$^+$.

 The second question concerns how large organic molecules are formed in
protostellar cores. In recent years, diverse organic molecules, including
methanol (CH$_3$OH), dimethyl ether (CH$_3$OCH$_3$), acetonitrile (CH$_3$CN),
and formic acid (HCOOH), have been detected towards low-mass protostars
\citep[][and references therein]{cec07}. They are still referred to as
"hot-core species", because it was once thought that they are only
characteristic of hot ($T\sim 200$ K) cores in high-mass star forming regions.
A large number of modeling studies
on hot-core chemistry show that sublimed formaldehyde (H$_2$CO) and CH$_3$OH are
transformed to other organic species by gas-phase reactions within a typical
timescale of $10^4-10^5$ yr \citep[e.g.][]{mh98}. In low-mass cores, however,
the time scale for the cloud material to cross the hot ($T\sim 100$ K) region
should be smaller than $10^4$ yr, considering the infall velocity and
temperature distribution in the core \citep{bot04a}. Furthermore, theoretical
calculations and laboratory experiments recently showed that  gas-phase
reactions are much less efficient in producing some hot-core species, such as
methyl formate (HCOOCH$_{3}$) and dimethyl ether, than
assumed in previous models \citep{hor04, gep06,gh06}.

Several model calculations have been performed on the chemistry that occurs
in low-mass protostellar cores. \citet{dot04} solved a detailed gas-phase reaction
network assuming a core model for IRAS 16293-2422, and succeeded in reproducing
many of the observed lines within 50 \%. The physical structure of the core;
i.e., its density and temperature distribution, was fixed with time. They assumed
gas-phase initial molecular abundances that  pertain to the high-mass hot-core
AFGL 2591. 
\citet{lbe04}, on the other hand, constructed
a core model that evolves from a cold hydrostatic sphere to a protostellar
core by combining a sequence of Bonnor-Ebert spheres with the inside-out collapse
model by \citet{shu77}. They solved a chemical reaction network that includes
gas-phase reactions along with adsorption and desorption of gas-phase/ice-mantle
species. The resulting
molecular distributions and line profiles are significantly different from
those of the static core models \citep{leb05}. The chemical network of
\citet{lbe04}, however, does not include large organic species.

In the present paper, we re-investigate molecular evolution in star-forming cores
with the partial goal of answering the two questions posed above.
We adopt a core model by \citet{mi00}; it accurately calculates the radial
distribution of temperature, which determines when and where the ice components
are sublimated. The model also enables us to follow molecular evolution smoothly
from a prestellar core to a protostellar core.
The chemistry includes both
gas-phase and grain-surface reactions according to \citet{gh06}; the surface
reactions in particular are important for producing organic molecules in a warming
environment.  Here, we report a solution of the reaction network following the
temporal variation of density and temperature in infalling shells to derive a
spatial distribution of molecules, including complex organic ones,
in a protostellar core.

\section{Model}
\subsection{Physical evolution of a star-forming core}

\begin{figure}
\plotone{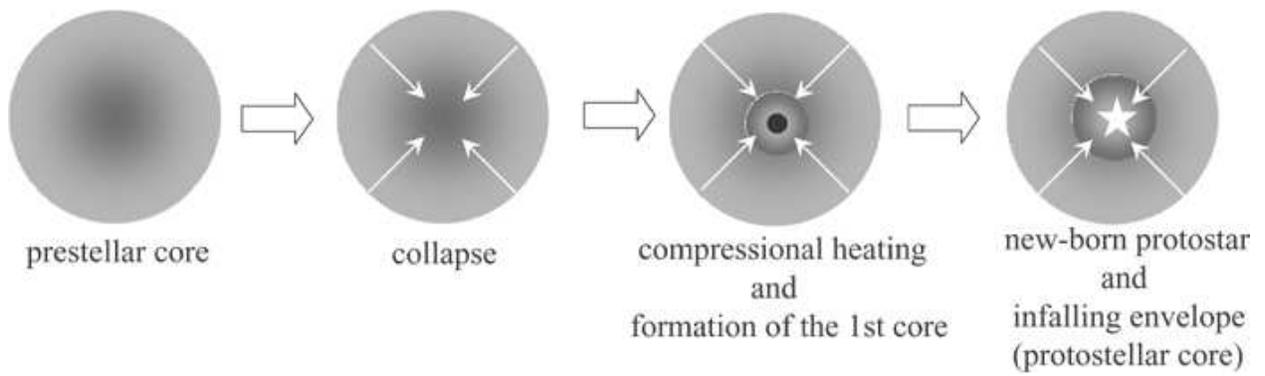}
\caption{Evolution of a star-forming core \label{schematic}}
\end{figure}

Figure \ref{schematic} schematically shows the evolution of a star-forming
core. As a model for this process, we adopted and partially reran the model
calculated by \citet{mi00}. These authors  solved the non-gray radiation
hydrodynamics to follow the core evolution from a dense starless (prestellar)
core to a protostellar core assuming spherical symmetry.
Conservation equations of mass, momentum, and energy were coupled with the
frequency-dependent radiation transfer, which was solved by the variable
Eddington factor method. \citet{mi00} calculated two models with different
initial conditions: a homogeneous core (one of uniform density)
and a hydrostatic core (a Bonnor-Ebert sphere). We chose the latter one in
the present work. 

Initially, the central density of the hydrostatic prestellar core is
$1.415 \times 10^{-19}$
g cm$^{-3}$, which corresponds to a number density of hydrogen nuclei
$n_{\rm H} \sim 6 \times 10^4$ cm$^{-3}$. The outer boundary is fixed at
$r=4\times 10^4$ AU, so that the total mass is $3.852 M_{\odot}$, which
exceeds the critical mass for gravitational instability. The temperature
in the core is initially $\sim 7$ K at the center and $\sim 8$ K at the outer
edge; the cooling by dust thermal emission is balanced with the heating by
cosmic rays, cosmic background radiation, and ambient stellar radiation.

In the original model by \citet{mi00}, the core starts contraction immediately.
In the present work, however, we assume that the core keeps its hydrostatic
structure for $1\times 10^6$ yr, implicitly assuming that turbulence supports
it.  This period sets the initial molecular abundances for the collapse stage. 
After $1\times 10^6$ yr, the core starts to contract. The contraction is almost
isothermal as long as the cooling
rate overwhelms the compressional heating, but eventually the latter dominates
and raises the temperature in the central region. Increasing gas pressure
decelerates the contraction and eventually makes the first hydrostatic core, known as
``the first core'',  at the center.
When the core center becomes as dense as $10^{-7}$ g cm$^{-3}$ and as hot as
$2000$ K, the hydrostatic core becomes unstable due to  H$_2$ dissociation
and starts to collapse again (the second collapse).  The central density increases rapidly,
and within a short period of time the dissociation degree approaches unity at the
center. Then the second collapse ceases and the second hydrostatic core; i.e.
the protostar, is formed. The protostar is surrounded by the infalling envelope, which we call the
protostellar core. 
After the onset of contraction, the initial prestellar core evolves to the
protostellar core in $2.5\times 10^5$ yr. After the birth of the protostar,
the model further follows the evolution for $9.3 \times 10^4$ yr, during which
the protostar grows by mass accretion from the envelope.
At each evolutionary stage, the model gives the total luminosity of the core and the
radial distribution of density, temperature, and infall velocity at
$r\gtrsim 10^{-4}$ AU. More detailed explanations of core evolution
can be found in \citet{mi98} and \citet{mi00}.

In the present paper, we report a re-analysis of their results in order to find the sublimation radii
of CO and large organic species at each evolutionary stage. We obtained electronic data
and the code of \citet{mi00} from the author (H. Masunaga), and re-ran the calculation
of the prestellar core phase to increase the number of time steps at which the core
structure is recorded. As for the protostellar phase, we used the existing data.
We are interested in the region outside 1 AU, which is of importance
for radio observations but is not discussed in detail by \citet{mi00}.
Based on this physical core model, we followed the temporal variation of density,
temperature, and visual extinction in infalling shells, in which the chemical
reaction network was solved \citep{aik01, aik05}.

\subsection{Chemical reaction network}

The chemical evolution was computed with the gas-grain model
developed in the OSU astrochemical group \citep{gh06}. The current version of
the model calculates the
abundance of 655 species (458 gas-phase and 197 grain-surface atoms and molecules) through
a total of 6309 reactions. The model follows the gas-phase as well as the
grain-surface chemistry using the rate equation method \citep[e.g.][]{hhl92,
rh00}. The gas-phase network is based on the osu.2005 database
available online (http://www.physics.ohio-state.edu/$\sim$eric/research.html)
but without the element fluorine, which has been recently added to osu.2005.
The model includes chemical reactions that can be important at
$T\lesssim 300$ K. At the final stage of our core model, the temperature
is somewhat higher than this limit at $r\lesssim$ 10 AU. In the following,
we mainly discuss the molecular evolution that occurs at $T\lesssim 300$ K.

The assumed elemental abundances, known as ``low-metal'' values because of
their strong depletions for elements heavier than oxygen,  are listed in
Table~\ref{abelem}. Initially the species are assumed to be in the form of
atoms or atomic ions except for hydrogen, which is entirely in its molecular form.
The cosmic-ray ionization rate $\zeta$ is set to $1.3 \times 10^{-17}$
s$^{-1}$.

The neutral species in the gas-phase stick to the grains upon collision with
a probability of 0.5. At high temperatures, this probability is much too high.
However, since evaporation is so rapid at such high temperatures, it is
unimportant what  assumption is made.
The grain-surface species can evaporate through thermal evaporation
and two non-thermal desorption processes. The binding energy of each molecular 
species on the grain surface is taken from \cite{gh06}.
The first non-thermal mechanism is
the action of a cosmic ray encountering a grain, which rapidly increases the
grain temperature and allows the evaporation of the species \citep{hh93}.
The second process, as recently developed quantitatively by
\citet{gpc06, gwh06},  is based on the assumption  that the
energy released by exothermic association reactions on grain surfaces can allow
the partial evaporation of the products \citep{wil68}.
In this process, the fraction of product evaporation is given by
\begin{equation}
f = \frac{a(1-E_D/E_{reac})^{s-1}}{1+a(1-E_D/E_{reac})^{s-1}},
\end{equation}
 where $E_{\rm D}$ is
the binding energy of the product, $E_{\rm reac}$ is the energy of formation in the 
reaction, and $s$ is the number of vibrational modes in the molecule/surface-bond system.
The parameter {\it a} is the ratio between the surface-molecule bond frequency and the
frequency at which the energy is lost to the surface. We assumed an $a$ value of 0.01
for all species based on \citet{gwh06}. With this value of $a$, less than 1\%
of water is evaporated during its formation on grains. Using molecular
dynamics calculations, \citet{ka06} showed that it is unlikely that this
fraction is larger than 1\%.
If it were, however, only the gas-phase
abundances of some large molecules such as CH$_3$OH would be increased in the
outer parts of the protostellar envelope.

Grain-surface species are dissociated by penetrating UV radiation and
cosmic-ray induced UV radiation.
We note that our model indirectly includes photodesorption via the second
non-thermal desorption process discussed above. For example, the photodissociation of
water ice produces OH radicals, a fraction of which recombine with hydrogen atoms to
re-generate water. Since we assume the partial desorption of the surface-reaction products, UV irradiation of water ice results in desorption of water.

Species can diffuse on the grain surfaces by
thermal hopping and react with each other when they meet. No quantum tunneling
was assumed even for hydrogen atoms based on recent experimental results
\citep[see][for discussion on this point]{gwh06}. Other details concerning the
model can be found in \citet{gh06} and \citet{gwh06}.

During the first $1\times 10^6$~yrs of the integration, we computed the chemical
abundances under static dense cloud conditions in order to obtain the initial
abundances for the collapse. The chemical evolution in infalling shells
was then computed as the shells experience
 temporal variations of density, temperature and visual extinction
$A_{\rm v}$. Although the original model of \citet{mi00} gives the temperatures
of gas and dust of various compositions separately, we assumed that
the dust temperature is equal to the gas temperature for simplicity.

We calculated the column density of hydrogen nuclei ($N_{\rm H}$)
from the core outer edge to each shell and converted it to $A_{\rm v}$ by
the formula $A_{\rm v}=N_{\rm H}/(1.59 \times 10^{21} {\rm cm}^{-2}$) mag.
Initially the visual extinction from the outer edge ($r=4\times 10^4$ AU)
to the core center is about 5.5 mag, and the shells we follow are located
at $\sim 1\times 10^4$ AU, where $A_{\rm v} \sim 1$ mag. Assuming that our
model core is embedded in ambient clouds, we added 3 mag to the visual
extinction obtained above, and ignored photodissociation of CO and H$_2$.
In \S 4, we also report molecular evolution in an isolated core,
in which the extra 3 mag is not added to $A_{\rm v}$

\section{Results}
\subsection{Physical evolution of the core and sublimation radius}

\begin{figure}
\plotone{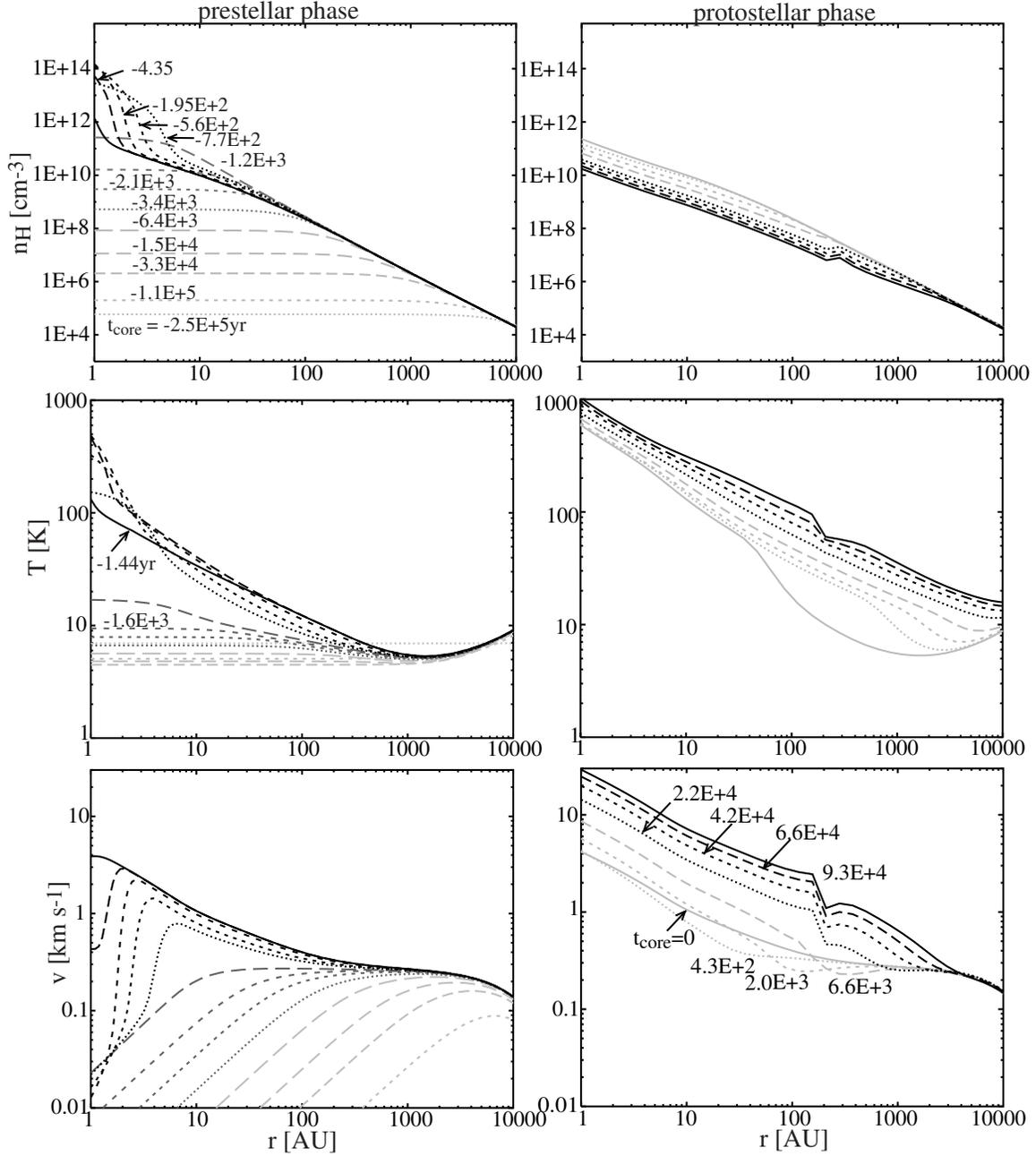}
\caption{Radial distributions of density, temperature and infall velocity in the prestellar core (left panels) and protostellar envelope (right panels). The times shown are relative to the birth of the second hydrostatic core, the protostar.  \label{MImodel}}
\end{figure}

Figure \ref{MImodel} shows the radial distributions of density, temperature and
infall velocity in the model core at assorted evolutionary stages.
We define the moment of the protostar formation as $t_{\rm core}=0$.
The left panels refer to the prestellar phase, which occurs before the birth of
the protostar, and the right panels to the protostellar phase. The distributions
are shown at different times in each panel.

We concentrate on the region outside 1 AU, which is of importance
for radio observations but is not discussed in detail by \citet{mi00}.
 At $t_{\rm core} < -1\times 10^3$ yr, more than
$1\times10^3$ yr before the formation of a protostar, the core is almost
isothermal and the core contraction is similar to the Larson-Penston (L-P)
collapse with a constant temperature \citep{lar69}.
At $t_{\rm core}\sim -1\times 10^3$ yr, the core center
starts to heat-up, and the core deviates from the L-P core. The increasing
pressure decelerates the contraction, and the first hydrostatic core, of radius $\sim 1$ AU,
is formed at $t_{\rm core}\sim -5.6\times 10^2$ yr (see the sharp velocity
gradient at $\sim 1$ AU in Figure \ref{MImodel}).
When the central density and the central temperature reach
$n_{\rm H}\sim$ several $10^{17}$ cm$^{-3}$ and $\sim 2000$ K, dissociation of
H$_2$ leads to the second collapse, which ceases in a short time scale
($\sim 1$ yr). In the protostellar stage, the density in the core envelope
decreases with time because of the accretion to the central star, while the
temperature and infall velocity increase.

\begin{figure}
\plotone{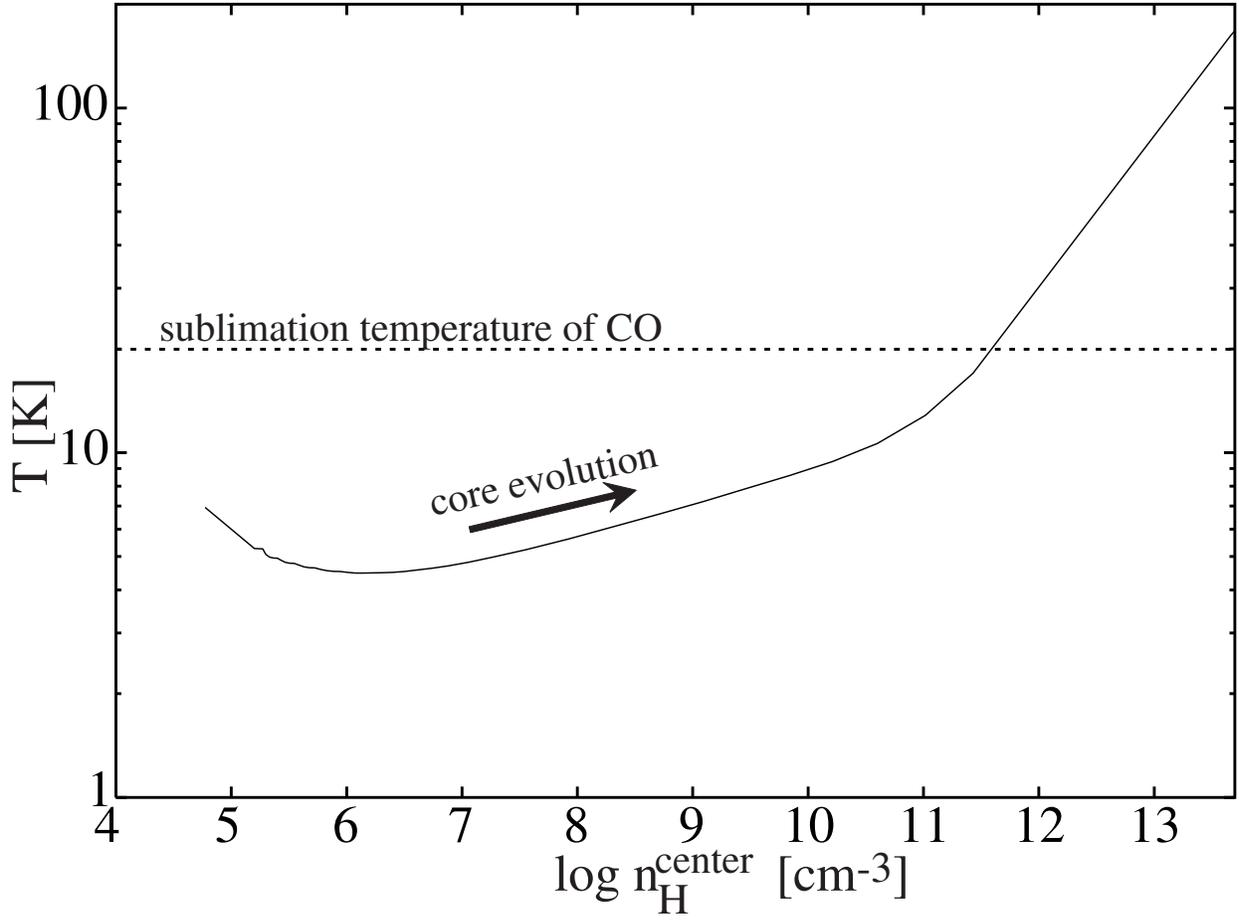}
\caption{Temporal variation of the central temperature of the prestellar core
plotted as a function of the central density.  \label{central_T}}
\end{figure}

In Figure \ref{central_T}, the temperature at the core center through 200 K is plotted as
a function of the central density $n_{\rm H}^{\rm center}$. The dashed line
indicates a temperature of 20 K, which is roughly the sublimation temperature
of CO, or the temperature at which the  rate coefficients of accretion and
thermal desorption for pure CO ice becomes equal.  In the prestellar
phase, the core temperature is determined by the balance between cooling by
thermal radiation and heating by cosmic rays, cosmic background radiation,
ambient stellar radiation, and compressional effects. The central temperature
for a time decreases to $\sim 5$ K, because the increasing column density of
the cloud prevents the penetration of optical photons.  The core temperature
starts to rise at $n_{\rm H}^{\rm center}\gtrsim 10^6$ cm$^{-3}$, when
compressional heating dominates radiative cooling. The temperature
increase is, however, rather moderate until the central density reaches
$\sim 10^{11}$ cm$^{-3}$. Shortly thereafter, CO sublimation starts when
the central density reaches a few $\times$ $10^{11}$ cm$^{-3}$. 
This occurs only several $\times$ $10^2$ yr before the
first-core formation and $\sim 10^3$ yr before the second-core formation.

\begin{figure}
\plotone{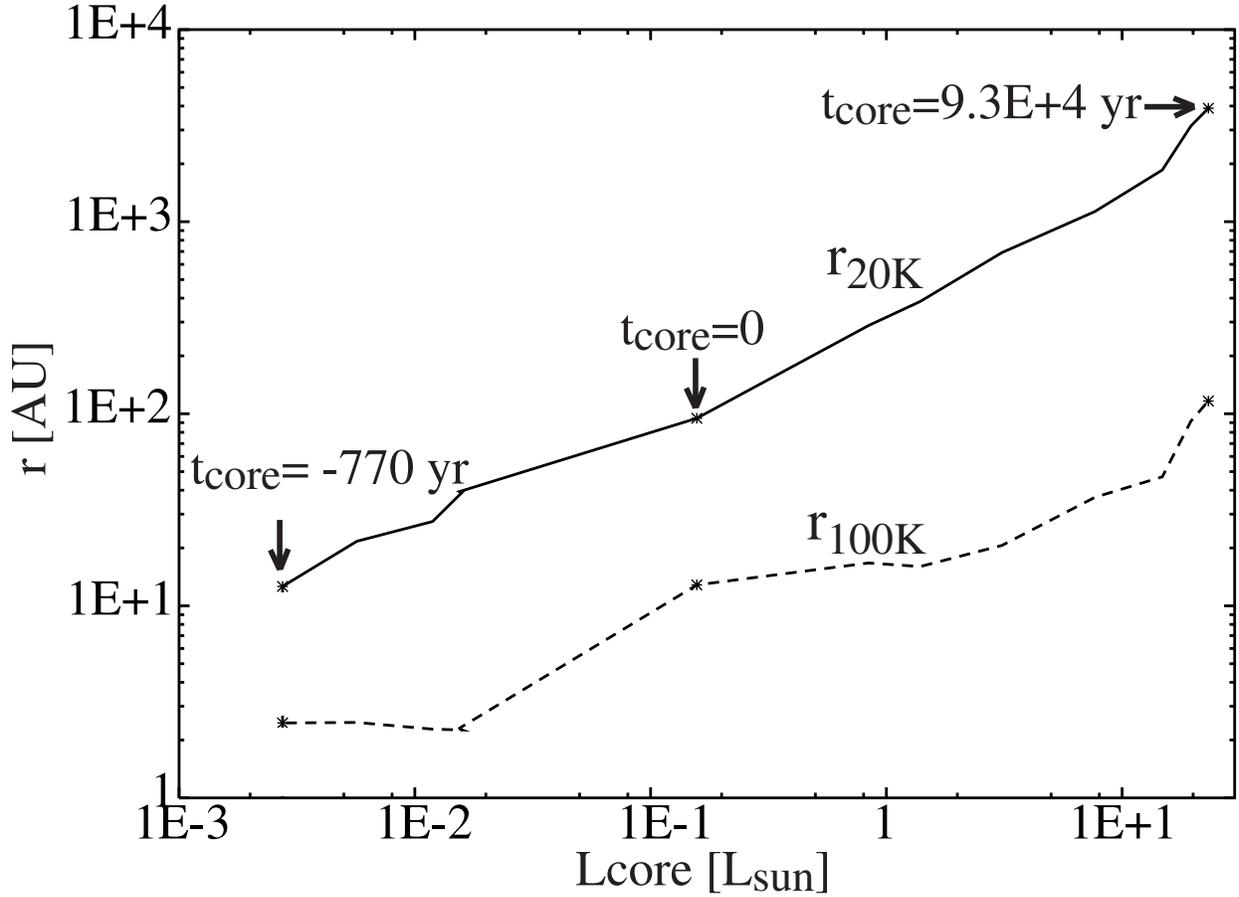}
\caption{Sublimation radius of CO (r$_{\rm 20K}$) and large organic species
(r$_{\rm 100K}$) as a function of total luminosity of the core. The solid and
dashed lines depict the radius inside of which the temperature is higher than
20 K and 100 K, respectively. \label{L_R20}}
\end{figure}

Figure \ref{L_R20} shows the temporal variation of two radii, inside of which
the temperature is higher than 20 K ($r_{\rm 20K}$) and 100 K ($r_{\rm 100K}$).
The former is roughly the sublimation temperatures of CO.  Desorption energies and thus sublimation temperatures vary among species,
but some large organic species such as CH$_3$OH and HCOOH, which are of
current interest, have sublimation temperatures of $\sim 100$ K.
The sublimation radii are plotted as
a function of total luminosity of the core instead of the core age, since the
former is an observable value. The labels with arrows depict the core age
$t_{\rm core}$. The plot starts about 770 yr before the second-core
formation, when the first core is about to be formed. As the first core grows
in density, the core luminosity and envelope temperatures increase.
At the moment of the
second-core formation ($t_{\rm core}=0$), $r_{\rm 20K}$ and $r_{\rm 100K}$ are
$\sim 100$ AU and $\sim 10$ AU, respectively. At the final stage of the model,
$t_{\rm core}=9.3\times 10^4$ yr, the total luminosity reaches $\sim 23
L_{\odot}$, and the sublimation radii are $r_{\rm 20K}\sim 3.9\times 10^3$ AU
and $r_{\rm 100K}\sim1.2\times 10^2$ AU.

\subsection{Physical conditions in infalling material}

Up to now, we have described the core model with physical parameters that are
given as functions of radius at each evolutionary stage, the so-called Eulerian
approach.  In order to calculate molecular evolution, however, it is more
convenient to follow the Lagrangian approach, in which we consider the temporal
variation of physical parameters in assorted collapsing shells, because this
temporal variation represents the conditions in which the chemical reaction
network is solved \citep{aik01, aik05}.  	

\begin{figure}
\plotone{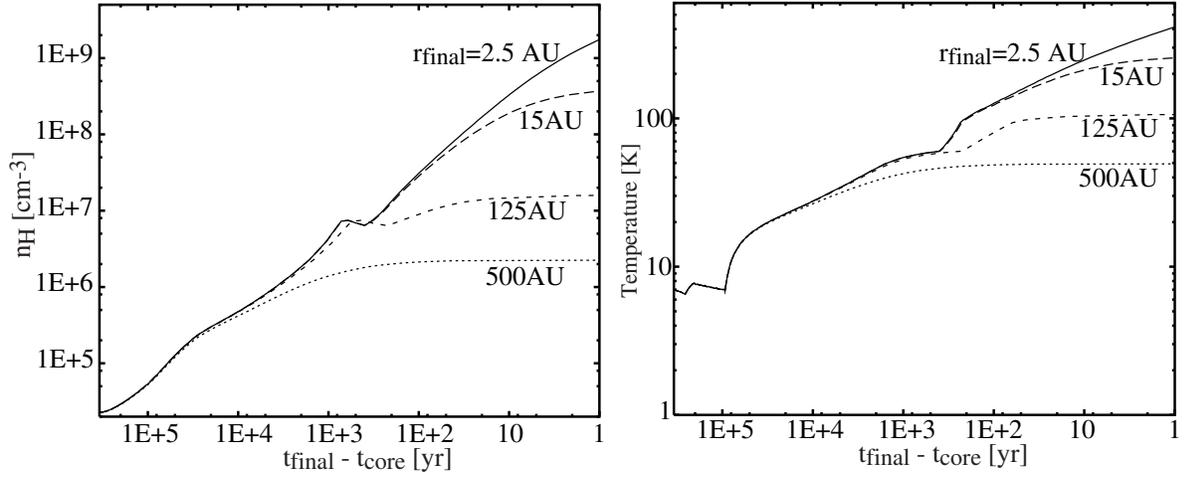}
\caption{Temporal variation of density and temperature in infalling shells
that reach $r=2.5, 15, 125$ and 500 AU at the final stage of our model
($t_{\rm core}=9.3\times 10^4$ yr). The horizontal axis is  the
logarithm of $t_{\rm final}-t_{\rm core}$.\label{fluid_parcel}}
\end{figure}

Figure \ref{fluid_parcel} shows the temporal variation of density and
temperature in shells that reach $r=2.5, 15, 125$ and 500 AU at the final
stage of our model ($t_{\rm core}=9.3\times 10^4$ yr $\equiv t_{\rm final}$). 
The infalling shells begin at $r\sim 1\times 10^4$ AU in the prestellar core, where
they stay for $1\times 10^6$ yr, a temporal period omitted in Figure
\ref{fluid_parcel}. After the core starts contraction, each shell migrates
inwards. In the prestellar stage, the variation of density and temperature is
not very significant; these shells are at $r\gtrsim 6\times 10^3$ AU,
where the infall velocity is small and the radial gradient of density and
temperature is moderate (Figure \ref{MImodel}). In the protostellar stage,
which occupies a large fraction of Figure \ref{fluid_parcel},
they migrate to inner radii where the infall velocity, temperature, and
density steeply increase inwards. Hence, the temporal variation of density
and temperature accelerate. In order to highlight the rapid variation near
and at the final stage, the horizontal axis in Figure \ref{fluid_parcel}
is set to be the logarithm of $t_{\rm final}-t_{\rm core}$.

A key parameter for the chemistry of hot-core species is the time scale of
the hot-core phase and/or, its predecessor, the warm-up phase.
In a classical hot-core chemical
model, in which the gas-phase reactions of sublimates are followed with a fixed
temperature of $\sim 200$ K, it takes $10^4-10^5$ yr to form new large organic
species. In other words, the hot-core phase of $10^4-10^5$ yr is needed to
account for the existence of large organic species \citep[e.g.][]{mh98}.
In more recent hot-core models, the temperature first increases with time
rather than remaining fixed; the warm-up phase is the period in which the
temperature rises from
10 K to $\sim 100$ K. Various reactions occur during the warm-up
phase, because many volatile species return to the gas phase, and because
the sublimation temperature varies with species \citep[e.g.][]{vw99}.
Grain-surface reactions between molecular radicals containing heavy elements
are also enhanced
in this temperature range \citep{gh06}. Considering the formation time scale
of stars, \citet{vit04} argued that the warm-up phase is longer in stars
with lower masses, and estimated the warm-up phase to be $10^6$ yr for
solar-type stars. Our model, in contrast, gives a warm-up phase of only
several $10^4$ yr; since the gas and dust are falling inwards towards the central star,
the timescale of the warm-up phase is determined not by the time scale of the star
formation, but by the size of the warm region divided by the infall velocity.
The timescale of the hot-core phase, in which the temperature of the infalling 
shells is $\sim 100-200$~K, is even shorter; the shell
reaching $r=2.5$ AU at $t_{\rm final}$ spends only 100~yr in the region of
$T\gtrsim 100$~K before it falls onto the central star. There would thus not be sufficient
time for the gas-phase reactions of sublimated species to form complex species.
We are dealing here
with the formation of a low-mass protostar, so the hot-core stage is more
aptly referred to as a ``hot corino'', and short time scales have previously
been discussed for such regions \citep{wakelam04,bot04a}. It should also be noted
that in our contracting core model, the central region with high temperatures
($T >$ several tens of K) is continuously fed by infalling materials, which are then lost to
the central star.

\subsection{Molecular evolution in infalling shells}

\begin{figure}
\epsscale{0.6}
\plotone{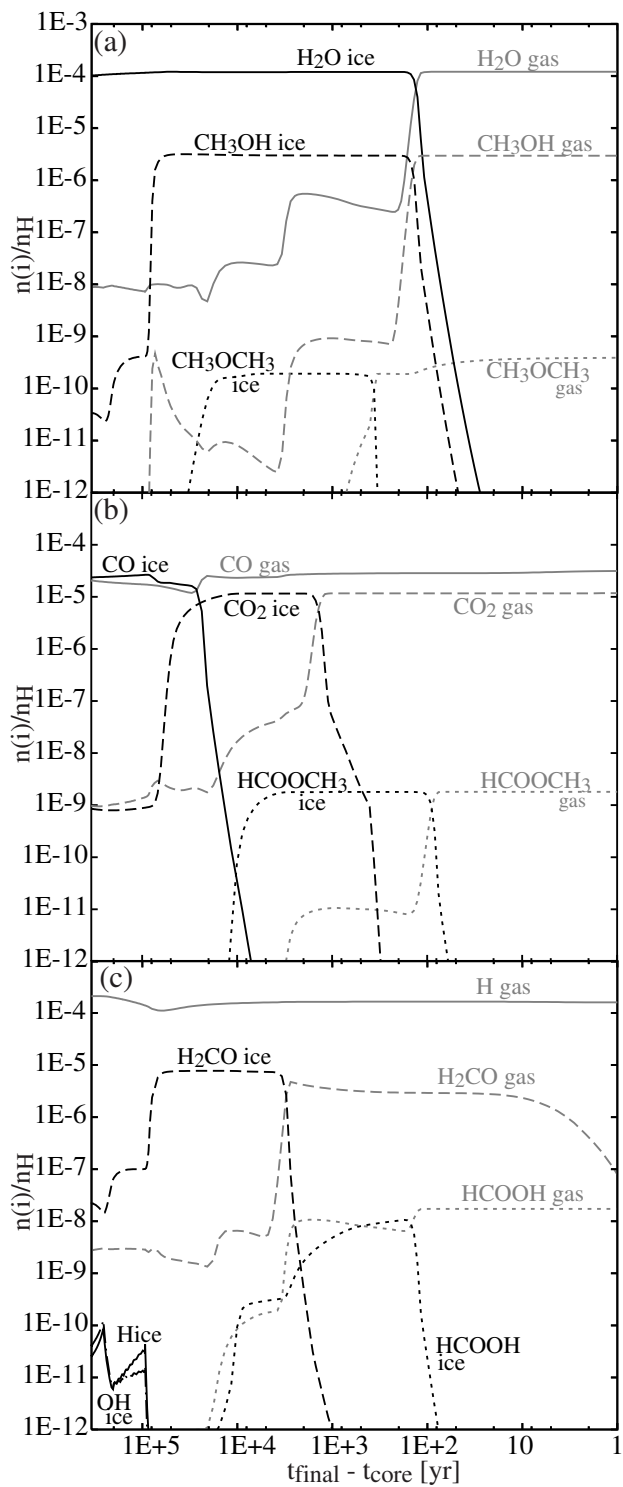}
\caption{Temporal variation of molecular abundances in the shell that reaches
$r=2.5$ AU at $t_{\rm final}$.
The horizontal axis is  the logarithm of $t_{\rm final}-t_{\rm core}$.
Black lines represent ice mantle species, while gray lines represent 
gas-phase species.
\label{evol_abun}}
\end{figure}

Figure \ref{evol_abun} shows the temporal variation of molecular abundances in
the infalling shell that reaches $r=2.5$ AU at $t_{\rm final}$. The
horizontal axis is again  the logarithm of
$t_{\rm final}-t_{\rm core}$. Black and gray lines represent species in the
ice mantle and in the gas phase, respectively. 
The chemical evolution at $t_{\rm final} -t_{\rm core}\le 4$ yr should be taken with
caution because the temperature in the shell exceeds 300 K, while our chemical
network is originally constructed for temperatures of $\lesssim 300$ K.
The species shown in Figure \ref{evol_abun}
are mostly (except CO and H) formed by grain-surface reactions.
They are desorbed into the gas phase when the shell temperature reaches the
sublimation temperature of each species; at this time the fractional abundance
of the surface species is simply transferred to a similar fractional abundance
in the gas. 

Water, the most abundant molecule with a heavy atom, is formed by the
hydrogenation of oxygen atoms on a grain surface, and is already abundant
when the contraction starts.  Methanol, on the other hand, takes longer to
reach a high abundance: it is formed by the hydrogenation of solid CO, which
occurs efficiently until the shell temperature reaches $\gtrsim 15$ K and
the H atom abundance in the ice mantle drops due to evaporation.
The fractional abundance of methanol exceeds 10$^{-6}$ at times within
10$^{5}$ yr of the final time. Then the abundance reaches a constant value;
CH$_3$OH is constantly destroyed by the cosmic-ray induced photodissociation
and re-formed by a low rate of hydrogenation.
Solid carbon dioxide is formed by O + HCO
$\rightarrow$ CO$_2$ + H and CO + OH $\rightarrow$ CO$_2$ + H.  It reaches
a fractional abundance of 10$^{-5}$ about 10$^{4}$ yr before the final time.

The large organic species are also formed mainly by grain-surface reactions.
Dimethyl ether (CH$_{3}$OCH$_{3}$)and methyl formate (HCOOCH$_3$) are
produced by the surface reactions of the radical H$_2$COH with CH$_3$
and HCO, respectively, where H$_2$COH is formed by hydrogenation of H$_2$CO.  
Formaldehyde, at this stage, is constantly formed by hydrogenation and destroyed
by a reaction with OH ice and cosmic-ray induced photodissociation.
Both CH$_{3}$OCH$_{3}$ and HCOOCH$_3$ reach terminal fractional abundances of 
$\approx 10^{-9}$.
Formic acid (HCOOH), on the other hand, is formed by the gas-phase reaction
of OH with sublimated H$_2$CO, after which it is adsorbed onto grains,
and remains in the ice mantle until the temperature gets high enough for
its sublimation.  Its asymptotic fractional abundance is slightly greater than
10$^{-8}$. Comparing Figure \ref{evol_abun} with Figure \ref{fluid_parcel},
we can see that these large organic species are mostly
produced at temperatures of $20-40$ K, rather than at $T\gtrsim 100$ K
where CH$_3$OH is sublimated into the gas phase \citep[see also][]{gh06}.
In this temperature range, heavy-element species such as CO, HCO and CH$_3$ can
diffuse on the grain surface and form complex molecules efficiently.

\subsection{Distribution of molecules in a protostellar core}
\begin{figure}
\epsscale{0.5}
\plotone{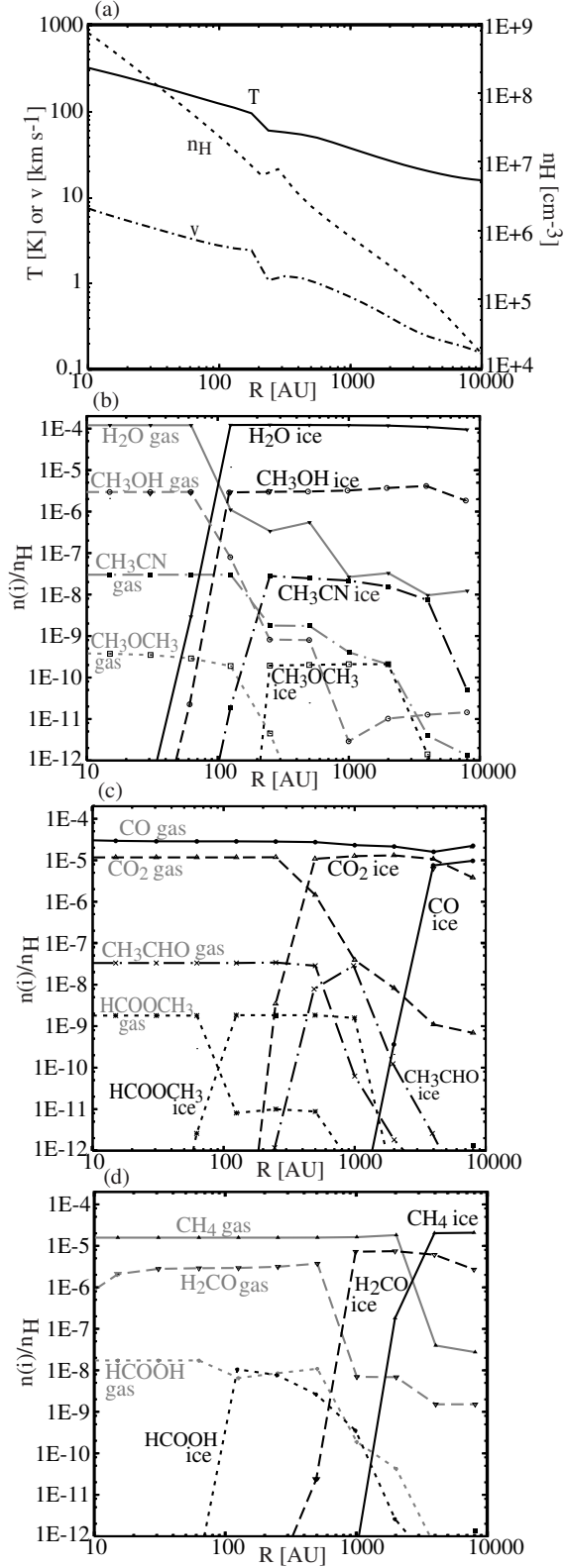}
\caption{Radial distribution of (a) physical parameters and
(b-d) molecules in the protostellar core at $t_{\rm final}$.
Black lines represent ice-mantle species, while gray lines
represent gas-phase species.
\label{dist}}
\end{figure}

In order to derive the spatial distribution of molecules in the protostellar
core, we have calculated molecular evolution in 13 shells, which reach
$r=2.5-8000$ AU at $t_{\rm final}$. Figure \ref{dist} shows the radial
distribution of (a) physical parameters and (b-d) molecular abundances
at $t_{\rm final}$.
 Abundances inside 10 AU are not shown for the following two reasons:
firstly they are very similar to the abundances at 10 AU, except that the
H$_2$CO abundance drops inwards, and secondly, the temperature in the inner
radius is somewhat higher than the temperature range originally
considered in our chemical model (\S 2.2).
Figure \ref{dist} (b-d) show that large organic molecules are abundant
in the gas phase in
the central region of the protostellar core. Since they are formed mostly
on grain-surfaces, rather than from gas-phase reactions among sublimates,
their gas-phase abundances sharply increase at radii corresponding to their
own sublimation temperatures and remain high at smaller radii. For example,
CH$_3$CHO extends to $\sim 500$ AU, while CH$_3$OH extends to $\sim 100$ AU.
One exception, again, is HCOOH, which extends to $\sim 500$ AU because of
its formation in the gas phase.

Large organic species are abundant in ice mantles at radii of $100-1000$ AU.
For radii $r\gtrsim 10^3$ AU,  the abundances of these species tend to
decrease sharply with increasing radius except for methanol. Water, another
hydrogen-rich, or saturated, species, remains nearly constant in abundance.
In the outer region ($r > 4000$~AU), a large fraction of carbon on the grain
surface is in the form of CH$_4$ \citep[cf.][]{gwh06}.

\section{Discussion}

\subsection{Physical structure of the core and sublimation radius}
We have re-analyzed and adopted the results of \citet{mi00} to derive the sublimation
temperatures of CO and large organic species,
and to investigate molecular evolution in a star-forming core. The chosen
conditions in the envelope can be different from those of other models because 
the temperature
distribution in the envelope depends not only on the evolutionary state
and mass of the central object (either the first or second core),
but also on the mass distribution in the envelope, which should depend
on the initial conditions.  Recently, \citet{omu07} investigated
the temperature distribution in the first-core envelope assuming 
mass distributions from the L-P model \citep{lar69} and Shu model
\citep{shu77} similarity solutions. The former has a more massive envelope
than the latter.
When the first core has a mass of $0.05 M_{\odot}$, for example, the
hydrogen number density at $r=10$ AU is several $10^{10}$ cm$^{-3}$ and
$\sim 10^9$ cm$^{-3}$
with the L-P model and the Shu model, respectively.  
The L-P model gives a core luminosity of $10^{-1} L_{\odot}$, and 
sublimation radii of $r_{\rm 20K} \sim 100$ AU and $r_{\rm 100 K}\sim 10$ AU,
while the Shu model yields a  core luminosity of $\sim 10^{-3} L_{\odot}$
and an $r_{\rm 20 K}$ of $\sim 20$ AU. The latter model does not exceed 100 K
at any radius.
Our core model is warmer than the Shu model but colder than the L-P model.

In the second core stage, on the other hand, the envelope structure can
deviate from spherical symmetry and be accompanied by a circumstellar disk.
Considering a typical angular velocity for  molecular cloud cores
\citep[$\sim 10^{-14}$ s$^{-1}$; e.g.,][]{ga85}, 
the centrifugal radius (i.e. the initial disk
radius) is $\sim 100$ AU. Our results at $r\gtrsim$ several hundred AU are thus
relatively robust, while the core structure would be significantly different from
our model at smaller radii. For example, the large organic species could be sublimated
by the accretion shock onto the forming disk rather than in the envelope, since
the centrifugal radius $\sim 100$ AU coincides with
the sublimation radius of large organic species in our model.

\subsection{Simple molecules}
\begin{figure}
\epsscale{0.5}
\plotone{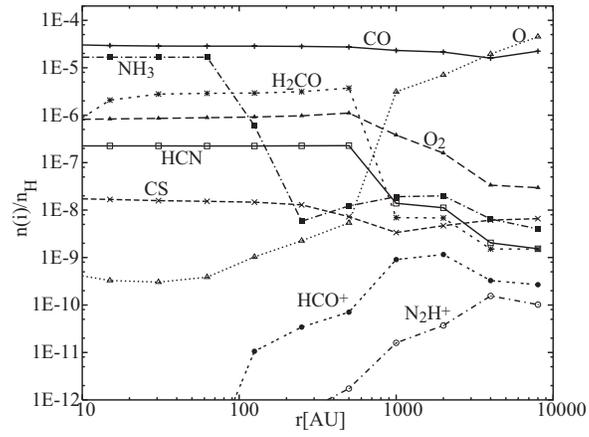}
\caption{Radial distribution of simple gaseous molecules at $t_{\rm final}$.
\label{cfLee}}
\end{figure}

\cite{lbe04} investigated the evolution of relatively simple molecules, such as HCN
and N$_2$H$^+$, in a star-forming core by combining a sequence of Bonnor-Ebert
spheres for the prestellar stage with the inside-out collapse model of \cite{shu77}
for the core after the first-core formation.
These simple molecules are often observed in star-forming cores
and are of importance as observational probes, while we mainly discussed  large
organic species in \S 3. Figure \ref{cfLee} shows the radial distribution of simple
molecules in our model at $t_{\rm final}$.

Comparison with \cite{lbe04} is not easy because there are many differences in 
the physical core models and chemical reaction networks. First,
the inside-out core model, which is adopted in \cite{lbe04}, yields lower temperatures
than our core model, as discussed above. Secondly, we adopt different binding energies
 for molecules
to the grain surface. While the difference is on the order of only a few hundred K for many species, the
binding energy of NH$_3$ is significantly higher in our model (5534 K) than in
\cite{lbe04} (1082 K); the latter value seems to originate from \cite{hh93}, in which
the hydrogen bonding of NH$_3$ is not taken into account.
Thirdly, grain-surface reactions, which are the main formation processes of saturated
species such as NH$_3$ and H$_2$CO in our model, are not included in \cite{lbe04}. Hence,
here we compare our results with \cite{lbe04} qualitatively rather than quantitatively.

A main conclusion
of \cite{lbe04} is that the molecular abundances vary significantly near and inside the
CO sublimation radius. For instance, N$_2$H$^+$ is destroyed by CO, and thus declines
steeply inwards across the CO sublimation radius, an effect which happens in our model as well.
\cite{lbe04} also found that some molecules, such as H$_2$CO, have their peak  abundance
at the sublimation radius; the gas-phase abundance first increases inward via
sublimation, but then decreases due to gas-phase reactions at smaller radii. Similar
phenomena can be seen in our model, but the spatial variation of the gas-phase 
abundances is more moderate than in \cite{lbe04}, for which we can think of a few
reasons. First, \cite{lbe04} calculated the molecular evolution in 512 shells,
while we calculated the chemistry in only 13 shells.
A larger number of shells are
needed to resolve abundance fluctuations on a smaller radial scale. Secondly, higher
infall speeds, in general, make the abundance distributions more uniform. Since the infall
speed is higher at inner radii in the infalling envelope, it would be natural that the
molecular abundances remain relatively uniform at $r\lesssim 100$ AU, which is not
calculated in \cite{lbe04}. Thirdly, our chemical network includes a much larger number
of species and reactions. \cite{lbe04} used a reduced reaction network with $\sim 80$
species and $\sim 800$ reactions to save computational time, while our
model includes 655 species and 6309 reactions. In a small reaction network, a sudden 
increase of a species (e.g., as caused by sublimation) can easily change the abundances of other
species though chemical reactions. In a large reaction network, on the other hand,
a larger number of reactions contribute to the formation and destruction of
each species, and thus the sudden abundance change of one species
does not necessararily propagate to
other species. Another noticeable difference in our results from those of  \cite{lbe04} is that HCO$^+$
decreases more steeply inwards at $r\lesssim 1000$ AU in our model;
H$_3$CO$^+$ and C$_3$H$_5^+$ are the dominant positive ions rather than HCO$^+$
at the inner radii.

The oxygen-bearing species atomic oxygen (O), molecular oxygen (O$_2$), and gaseous H$_2$O are also of special interest
because O  is very reactive and because
the  two molecules have been intensively 
observed by {\it the Submillimeter Wave Astronomy Satellite (SWAS)} and {\it Odin} 
Satellite in recent years.  Our model predicts that O  reaches
an abundance of  $4\times 10^{-5}$ relative to hydrogen nuclei at $r=8000$ AU,
while it steeply decreases inwards from $\sim 1000$ AU to 100 AU.  Even at $r=8000$ AU, however,
H$_2$O ice is the most abundant O-bearing species (Figure \ref{dist}).
\cite{ber00} summarized the {\it SWAS} observations of cold molecular clouds by stating that
the fractional abundance of O$_2$ lies under $10^{-6}$ and that of gaseous water lies in the range
$10^{-9}$ to  a few $\times$ $10^{-8}$. Molecular oxygen has recently been
detected towards $\rho$ Oph by Odin with an abundance of $5\times 10^{-8}$
relative to hydrogen \citep{lar07}. These abundances are consistent with our
predictions for the 
outermost radius $r=8000$ AU, where the density $n_{\rm H}\sim 10^4$ cm$^{-3}$ and
temperature $T<20$ K are similar to those in molecular clouds.

\subsection{Dependence on visual extinction at the core edge}

So far we have assumed that the model core is embedded in ambient clouds of
$A_{\rm v}= 3$ mag. In reality, some cores (e.g. Bok globules) are
isolated, while others are embedded in clouds. Isolated cores are directly irradiated
by interstellar UV radiation, which causes photo-reactions (photodissociation and ionization) in the gas-phase and ice
mantles \citep[e.g.][]{lee96,rh01}.
In order to evaluate the effect of ambient UV radiation,
we recalculated molecular abundances
in a core that is directly irradiated by interstellar UV radiation; i.e. $A_{\rm v}=0$
mag at the outer edge of the core ($r=4\times 10^4$ AU).
The photodissociation rates of H$_2$ and CO were calculated
by following \citet{lee96}, which gives the shielding factors as a function
of $A_{\rm v}$ and column densities of CO and H$_2$ in the outer radii.
The outermost shell for which we calculate molecular evolution is initially
located at $r\sim 1.4\times 10^4$ AU and declines to $r=8000$ AU at
$t_{\rm final}$. Column densities of CO and H$_2$ outside of this shell were
estimated by assuming $n$(CO)/$n_{\rm H}=5\times 10^{-5}$ and
$n$(H$_2$)/$n_{\rm H}=0.5$.

\begin{figure}
\plotone{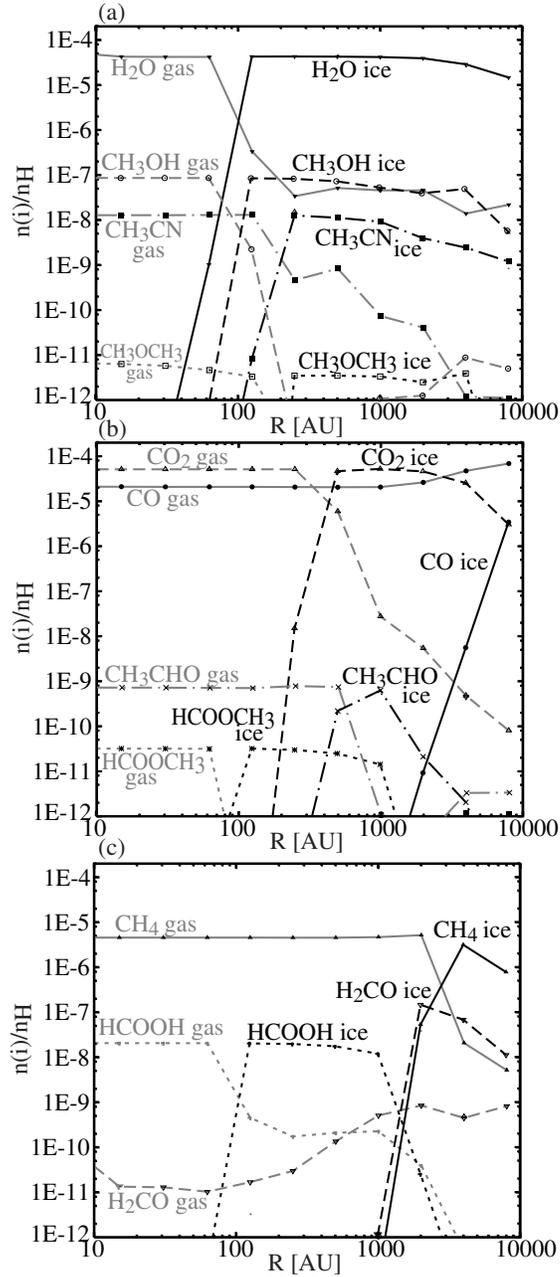}
\caption{Radial distribution of molecules in the protostellar core, as in Figure 7
(b-d). The core is assumed to be directly irradiated by the interstellar UV radiation.
\label{av0}}
\end{figure}

Figure \ref{av0} shows the resultant distribution of molecular abundances
in a protostellar core at $t_{\rm final}$. Compared with the embedded core
model (Figure \ref{dist}), the fractional abundances of CH$_3$OH and H$_2$CO
are lower by more than one order of magnitude, while the CO$_2$ abundance is
higher. When the shells are still at $r\gtrsim$ several thousand AU and have
relatively low $A_{\rm v}$ ($\lesssim 4$ mag), the photodissociation of
H$_2$O ice efficiently produces OH, which reacts with CO
to produce CO$_2$ ice. Methanol in the ice mantle is dissociated to H$_2$CO,
which is further dissociated to CO. Species such as CH$_3$CHO,
HCOOCH$_3$ and CH$_3$OCH$_3$ are also less abundant in the isolated model,
because their formation path includes H$_2$CO in the ice mantle. Formic acid
in the gas phase extends only up to $\sim 100$ AU, while it extends to several
hundred AU in the embedded model. In the isolated model, it is formed mainly
by OH + HCO on the grain surface.
Our results may indicate that molecular abundances in hot cores and corinos
depend on whether the core is isolated or embedded in clouds.

The importance of photo-reactions on ice mantle abundances has also been
investigated in  a number of laboratory experiments.   For example, \citet{wk02} and \citet{wm07} found that carbon dioxide is efficiently 
produced by UV irradiation
on  a binary ice mixture of H$_2$O and CO, a result that is consistent with our model.
However, \cite{wm07}  found that the UV irradiation also produces CH$_3$OH with a
relative abundance of $n$(CH$_3$OH)/$n$(CO) $\sim 10$ \% in the ice mixture. 
Comparison of our model with
their experiment is not straightforward because of  differences in temperature and
included reactions,  but we
may have underestimated the CH$_3$OH ice abundance in the irradiated core model.
The discrepancy can arise from the H atom desorption rate in our model. We calculated that
UV radiation penetrates into the ice mantle and dissociates H$_2$O to produce
H atoms embedded in ice. Although we do not discriminate between H atoms
on the ice surface and those embedded in the ice mantle, the embedded H atoms
in reality would have a lower desorption rate and a better chance of reacting with
neighboring CO, a reaction that initiates the formation of methanol in the mantles. Discrimination between surface and embedded hydrogen atoms should be included in future work.

\subsection{Comparison with observation}

Our model results can be compared with observational results of low-mass
protostars. Comparison of the physical structure of the core has  already been
discussed in detail by \citet{mi00}.  Here we concentrate on molecular
abundances. First, we compare molecules in ice mantles. The observation of ices
towards the low-mass protostar Elias 29 is summarized in \citet{es00};
the abundances of CO, CO$_2$, CH$_3$OH and CH$_4$ relative to water ice are 5.6 \%,
22 \%, $< 4$ \%, and $< 1.6$ \%, respectively. On the other hand, \citet{ppp03} detected 
a high abundance (15-25 \% relative to water ice) of CH$_3$OH ice towards 3 low-mass
protostars among 40 observed protostars. Although the observation of ice
features is difficult, it is probable that the composition
of ice mantles depends on their environment and the history of the observed
regions.
From a theoretical point of view, the abundances of molecules on grains and
their fractional abundances compared with H$_2$O ice depend on time and
radius from the protostar. In addition, an embedded core and an isolated core
lead to significantly different calculated abundances for solid CH$_3$OH,
H$_2$CO and CO$_2$.

Table~\ref{solid_theo} lists ratios of ice species with respect to water ice at $t_{\rm final}$ for local abundances at radii of 1000 AU and 8000 AU and for column densities towards the core center.
Since most of the core material exists at $r\le 8000$ AU along the line of sight,
and since the ices are present at $r>10$ AU, the column density is calculated by 
integrating the number density of ice species from 10 AU to 8000 AU.
It is interesting to note that regardless of the distance from the protostar,
the isolated core model gives smaller surface abundances of CH$_3$OH and
H$_2$CO and a higher surface abundance of CO$_2$ than the embedded model.
In both models, the surface abundance of CO at
1000 AU is much smaller than observed, but it increases towards larger radii
up to 10 and 20\% in the embedded and isolated models respectively. Comparison with the observations is best done using our column density ratios, which are in reasonable agreement with Elias 29 for both models.
Considering the variation among cores, our models show reasonable agreement
with observation.  The disagreement with CO doubtless results from our
assumption concerning desorption rates. In the present work, we use a desorption
energy for each species mainly referring to laboratory experiments on pure
ice sublimation or theoretical estimates that sum up the van der Waals forces
between adsorbed atoms and grain surface. But in reality, interstellar ice is a
mixture, and hence volatile species can be entrapped by less volatile
species; recent temperature-programmed desorption results show that in a
mixture rich in water ice, much CO desorbs at considerably higher temperatures
\citep{cdf03}.

Table \ref{obs} lists the gas-phase abundances of large molecules towards the
low-mass protostar IRAS 16293-2422 and in the central region ($r=30.6$ AU)
of our core model. The physical and chemical structures of IRAS 16293-2422
have been intensively studied \citep{cec00a,cec00b, sch02,cau03,
cha05}. The physical parameters derived by the model at $t_{\rm final}$
($\sim 23 L_{\odot}$ and $r_{\rm 100K}\sim1.2\times 10^2$ AU) are very close to
the ones of IRAS16293-2422; the observed luminosity of the source is 27
$L_{\odot}$ \citep{wal86} and the physical structure has been constrained by
multi-line analysis of H$_2$O and H$_2$CO observations through a detailed
radiative transfer code \citep{cec00b}.
It should be noted that the emission lines of large organic molecules observed
in this source are not spatially resolved, except for a few lines investigated
by interferometric observations \citep{bot04b, kua04, rem06}.
The density and temperature of the gas should vary both within the beam and
along the line of sight. Hence the derived molecular abundances vary significantly
depending on the assumptions concerning the physical structure of the core and  the
emitting region, which explains the difference in abundances
determined by different investigators (see Table~ \ref{obs}).

Although it is not obvious which core model, embedded or isolated, should
be compared with IRAS16293-2422, we would prefer
the embedded core model. While IRAS16293-2422 is in the Ophiuchus molecular cloud,
which harbors several UV sources in the form of OB stars, the $^{13}$CO and C$^{18}$O
observations indicate that the core is embedded in dense gas \citep{lor89,tac00}.
The high molecular D/H ratios observed towards IRAS 16293-2422 and its neighboring
starless core IRAS16293E \citep{cec07,vas04} indicate that these cores have been
very cold and thus well-shielded from the ambient stellar radiation. 
Considering the uncertainties in observationally-estimated molecular abundances, HCOOCH$_3$, HCOOH, and CH$_3$CN in our embedded
core model show reasonable agreement with the observations. The embedded model,
however, underestimates CH$_3$OCH$_3$ and overestimates H$_2$CO and
CH$_3$OH. The isolated core model, on the other hand, reproduces observed
abundances of CH$_3$OH, HCOOH and CH$_3$CN, but strongly underestimates H$_2$CO,
HCOOCH$_3$, and CH$_3$OCH$_3$.  

We can think of several possible solutions to improve the agreement with
observation. First, it is noteworthy that the gaseous CH$_3$OH abundance
estimated in IRAS 16293-2422 is much lower than the abundance of CH$_3$OH ice
($n$(CH$_3$OH ice)/$n_{\rm H} \sim 10^{-5}$, assuming n(H$_2$O ice)/
$n_{\rm H}\sim 10^{-4}$) detected by \citet{ppp03} towards some low-mass
protostars. We may have missed or underestimated the reactions which
transform CH$_3$OH to other large organic species. Secondly, the abundances
of large organic species in the central region can vary with time.
In the present work, we have concentrated on the distribution of molecules only at
$t_{\rm final}$. Shells that reach the central region at different times
should experience different temporal variations of physical conditions.
Some shells may experience longer periods at $T\sim 20-40$ K, where large
organic species start to be efficiently formed. This possibility will be pursued
in a future work. Thirdly, core models with rotation could produce higher
abundances of large organic species; because of the centrifugal force,
core material migrates more slowly and stays longer in the temperature
range preferable for the formation of large organic molecules
(e.g. in a forming disk).

Because of the beam size of the current radio observations, little is known
concerning the spatial distribution of the large organic species; they are mostly
confined within a few arc seconds from the core center \citep[e.g.][]{kua04}.
But there are exceptions; \citet{rem06} found that HCOOH and HCOOCH$_3$ show
extended emission of $\sim 5$ arcsec. Our embedded core model reproduces the
extended emission of HCOOH, while HCOOCH$_3$ is confined to $r<100$ AU.

%

\subsection{Carbon chains in a protostellar core}

Recently, \cite{sshk07} detected strong emission lines of carbon
chain species such as C$_4$H and C$_4$H$_2$ towards the low-mass protostar L1527,
which is considered to be in a transient phase from class 0 to class I.
In general, carbon-chain species are associated with  the early
stages of a cold cloud core when the dominant form of carbon changes from
atomic carbon to CO. Hence, the existence of carbon-chain species towards L1527
is a surprise.

\begin{figure}
\epsscale{1.0}
\plotone{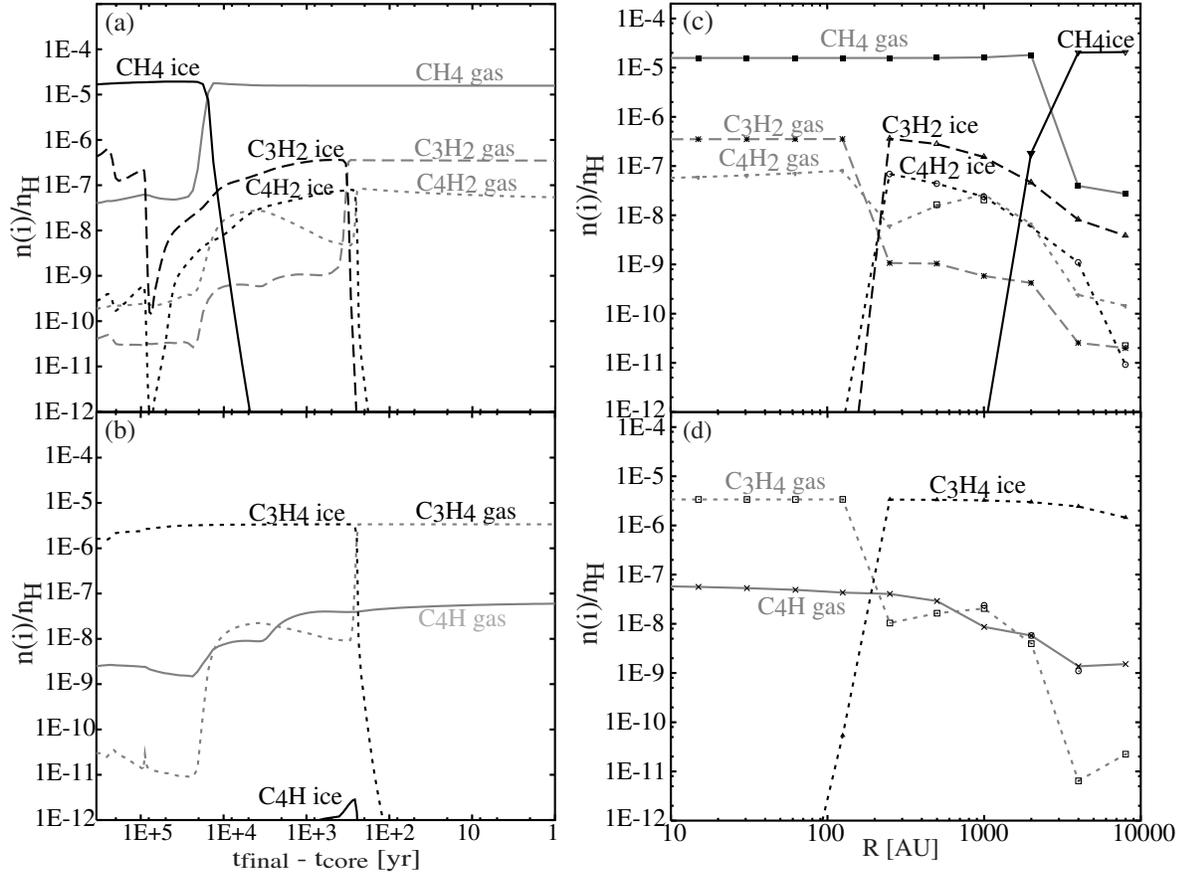}
\caption{(a-b) Temporal variation of carbon-chain abundances in the shell
that reaches $r=2.5$ AU at $t_{\rm final}$, as in Figure 6.
(c-d) Radial distribution of carbon-chain species in the protostellar core,
as in Figure 7. 
\label{carbon_chain}}
\end{figure}

Figure \ref{carbon_chain} (a-b) shows the temporal variation of CH$_4$ and carbon-chain
abundances in the shell that reaches $r=2.5$ AU at $t_{\rm final}$.  When the
core starts to contract, methane (CH$_4$) and C$_3$H$_4$ are already abundant
in ice mantles.  Methane has been formed by the hydrogenation of carbon on
grain surfaces, while C$_3$H$_4$ has been formed by a combination of gas-phase reactions
(to form unsaturated carbon chains such as C$_{3}$H and C$_{3}$H$_{2}$) and grain-surface reactions (to hydrogenate them). When the
CH$_4$ sublimates, some fraction reacts with C$^+$ to form
C$_2$H$_3^+$, which is a precursor for ion-molecule reactions that lead to the production of larger unsaturated hydrocarbons. While gas-phase reactions tend to
make the carbon chains longer,  adsorbed species
experience hydrogenation and dissociation by cosmic-ray induced UV radiation.
Figure \ref{carbon_chain} (c-d) shows the distribution of carbon-chain species in our
embedded core model. Methane and
C$_3$H$_4$ are abundant in the ice mantle even at $r\gtrsim 10^3$ AU, while
other hydrocarbons are abundant at a few 100 AU $\lesssim r \lesssim 10^3$ AU.
Most of the carbon-chain species  desorb at $r\lesssim$
a few 100 AU.  

In summary, our model indicates that carbon-chain species can be re-generated
in the protostellar core by the combination of gas-phase reactions and
grain-surface reactions. It should be noted, however, that large oxygen-containing organic species
are not detected in L1527 \citep{sak07}, while they are abundant in the central
region of our model. In future work, the temporal variation of the molecular distribution
in model cores should be investigated to see if at certain evolutionary stages
carbon-chain species are abundant while large organic species are not.  For example, in their recent study of the chemistry of cold cores, \cite{gwh06} found that generation of gas-phase hydrocarbons from precursor surface methane occurs at very late times, after the abundance of surface methanol has essentially vanished.
More detailed  observations are also needed for a quantitative comparison
with models. While our model predicts the column density of C$_4$H to be
$2\times 10^{16}$ cm$^{-2}$ towards the central star,
the observation gives $2\times 10^{14}$ cm$^{-2}$. The actual column density
towards the central star can, however, be larger, because the emission is averaged
within the beam (a few 10 arcsec) in the current observation.
Methane ice, the precursor of the carbon-chain species in our model, is as abundant
as 23 \% relative to water ice at the outer edge of our core model. A deep integration of
ice features towards field stars is needed to constrain the CH$_4$ ice abundances
at the outer edge of the core and/or quiescent clouds, while the column
density ratio of CH$_4$ ice to H$_2$O ice (1 \%, see Table 2) in our model is
consistent with the
observation towards YSO's (\S 4.3), because of the relatively large sublimation
radius of CH$_4$.

\section{Summary}

We have adopted the one-dimensional radiation-hydrodynamical core model of
\citet{mi00} to investigate  molecular evolution in a low-mass
star-forming core.
The physical structure (density, temperature and infall velocity) of the core
is given as a function of time and radius (distance from the core center).
The temporal variation of density, temperature and visual extinction is calculated
for assorted infalling shells. The temporal variation of molecular abundances
in these infalling shells and the radial distribution of molecules in the
protostellar core are calculated by solving a chemical reaction network
that couples gas-phase and grain-surface chemistry. 

In the prestellar phases of star formation, species such as CO, which contain heavy
elements,
are depleted onto grain surfaces due to low temperatures in the vicinity of 10 K and below.
As the contraction proceeds, the compressional heating
overwhelms the radiative cooling.  When the central density reaches $\sim
10^{11}$ cm$^{-3}$, the temperature at the core center starts to rise sharply
and soon reaches the sublimation temperature of CO ($\sim 20$ K). The first
hydrostatic core is formed shortly (in several $10^2$ yr) thereafter.
When the protostar is born, the CO sublimation radius
extends to 100 AU, and the temperature at $r\lesssim 10$ AU is higher than
100 K, at which some large organic species start to evaporate.

We investigated the radial distribution of molecules at $9.3\times 10^4$ yr
after the birth of a protostar, when the temperature is higher than 20 K and
100 K at $\lesssim 3.9\times 10^3$ AU and $\lesssim 100$ AU, respectively. The
time taken by the infalling shells to warm up from 10 K to 100 K is important for the
production of large molecules. We found that this warm-up phase occurs for only
several $10^4$~yr in our dynamical model. Once the shells enter the region
where $T\gtrsim 100$ K, they fall into the central protostar in the very short
time of $\sim 10^2$ yr.
Large organic molecules such as CH$_3$OH, HCOOCH$_3$ and CH$_3$OCH$_3$ are
formed on grain surfaces at temperatures of $20-40$ K and subsequently released
into the gas phase by thermal evaporation. In our model, only the abundance of
HCOOH seems to be influenced by gas-phase chemistry. As a consequence, the
radius at which the gas-phase abundance of an organic molecule typically increases
strongly in the envelope corresponds to its sublimation radius. 
Our model also indicates that carbon-chain species can be formed
by a combination of gas-phase reactions and grain-surface reactions
after the sublimation of CH$_4$.

We compared  molecular abundances in an isolated core model,
which is exposed to the ambient UV field,  and an embedded core model.
We found that the photo-reactions in ice mantles are important in determining the
ice abundances and thus the molecular abundances in hot corinos. The molecule CO$_2$, in particular,
is enhanced in the isolated model, while methanol and formaldehyde are
more abundant in the embedded model. Our model shows acceptable agreement
with gas-phase observations of the hot corino IRAS 16293-2422
and ice-phase observations towards the low-mass protostar Elias 29,
considering uncertainties and
variations in observed abundances. A more detailed comparison, including a
radiative transfer treatment to directly compare with the observed spectra,
is desirable.

\acknowledgments

This work was supported by a Grant-in-Aid for Scientific Research (17039008,
18026006)
and by  ``The 21st Century COE Program of Origin and Evolution of
Planetary Systems" of the Ministry of Education, Culture, Sports, Science
and Technology of Japan (MEXT).
E. H. thanks the National Science Foundation for support of his research program
in astrochemistry.

\clearpage
\begin{table}
\caption{Elemental abundances with respect to H.}
\begin{center}
\begin{tabular}{|c|c|}
\hline
\hline
Element & Abundance \\
He & 9.75(-1)\tablenotemark{a} \\
N & 2.47(-5)\\
O & 1.80(-4)\\
C$^+$ & 7.86(-5)\\
S$^+$ & 9.14(-8)\\
Si$^+$ & 2.74(-9)\\
Fe$^+$ & 2.74(-9)\\
Na$^+$ & 2.25(-9)\\
Mg$^+$ & 1.09(-8)\\
P$^+$ & 2.16(-10)\\
Cl$^+$ & 1.00(-9)\\
\hline
\end{tabular}
\end{center}
\tablenotetext{a}{$a(-b)$ means $a\times 10^{-b}$.}
\label{abelem}
\end{table}%

\begin{table}
\caption{Percentage abundances of ice-mantle species compared with H$_2$O ice in
model cores at $t_{\rm final}$. For each core model, ice abundance at radius of 1000 AU
and 8000 AU are listed. Ratios of column densities integrated towards the core
center are also listed as "column".}
\begin{center}
\begin{tabular}{l c c c c c c}
\hline
Species & \multicolumn{3}{c}{Embedded core} &
\multicolumn{3}{c}{Isolated core} \\
        & 1000 AU\tablenotemark{a} & 8000 AU& column &1000 AU & 8000 AU &column \\
\hline
CO     & $7\times 10^{-11}$ & 10 & 0.4 & $8\times 10^{-9}$ & 23 & 0.04 \\
CO$_2$ & 10               & 4 & 4.2 & 125            & 21 & 45\\
H$_2$CO &  6              & 3 & 1.5 & $2\times 10^{-6}$ & 0.08 & 0.03 \\
CH$_3$OH &  3             & 2  & 2.7 & 0.1             & 0.04 & 0.2 \\
CH$_4$ & $3\times 10^{-9}$ & 23 & 1 & $2\times 10^{-9}$ & 5 & 0.4 \\
\hline
\end{tabular}
\end{center}
\tablenotetext{a}{radii}
\label{solid_theo}
\end{table}%


\begin{table}
\caption{Gas-phase molecular abundances in IRAS 16293-2422 and model results.}
\begin{center}
\begin{tabular}{l c c c}
\hline
Species & IRAS 16293-2422 & \multicolumn{2}{c}{model}\tablenotemark{a}\\
        &                 & embedded & isolated \\
\hline
H$_2$CO & 1.0(-7)\tablenotemark{b}, 1.1(-7)\tablenotemark{c} & 2.8(-6) & 1.3(-11)\\
CH$_3$OH & 1.0(-7)\tablenotemark{d}, 9.4(-8)\tablenotemark{c}& 3.0(-6) & 8.5(-8) \\
HCOOCH$_3$ &  2.5-5.5(-7)\tablenotemark{e}, 2.6-4.3(-9)\tablenotemark{f},
              $>$ 1.2(-8)\tablenotemark{g} & 1.8(-9) & 3.2(-11)\\
HCOOH   & 6.2(-8)\tablenotemark{e}, 2.5(-9)\tablenotemark{g} & 1.7(-8) & 2.1(-8)\\
CH$_3$OCH$_3$ & 2.4(-7)\tablenotemark{e}, 7.6(-8)\tablenotemark{c} & 3.5(-10) & 5.8(-12)\\
CH$_3$CN & 1.0(-8)\tablenotemark{e}, 7.5(-9)\tablenotemark{h} & 3.0(-8) & 1.3(-8)\\
\hline
\end{tabular}
\end{center}
\tablenotetext{a}{Gas-phase abundances at $r=30.6$ AU are listed, because the abundances
are mostly constant at $r\lesssim 100$ AU.}
\tablenotetext{b}{\citet{mar04}}
\tablenotetext{c}{\citet{cha05}}
\tablenotetext{d}{\citet{mar05}}
\tablenotetext{e}{\citet{cau03}}
\tablenotetext{f}{\citet{kua04}}
\tablenotetext{g}{\citet{rem06}}
\tablenotetext{h}{\citet{sjv02}}
\label{obs}
\end{table}%

\clearpage

\clearpage


\begin{thebibliography}{}


\bibitem[Aikawa et al.~(2001)]{aik01} Aikawa, Y., Ohashi, N., Inutsuka,
S.-I., Herbst, E., \& Takakuwa, S. 2001, ApJ, 552, 639

\bibitem[Aikawa et al.~(2005)]{aik05} Aikawa, Y., Herbst, E., Roberts, H.,
\& Caselli, P. 2005, ApJ, 620, 330


\bibitem[Allen \& Robinson (1977)]{ar77}
Allen, M. \& Robinson, G. W. 1997, ApJ, 212, 396


\bibitem[Bergin \& Langer (1997)]{bl97}
Bergin, E. A., \& Langer, W. D. 1997, ApJ, 486, 316

\bibitem[Bergin et al.~(2000)]{ber00}
Bergin, E.A., et al. 2000, ApJ, 539, L129

\bibitem[Bergin \& Tafalla (2006)]{bt06} Bergin, E.A. \& Tafalla, M.
2006, ARA\&A, in press


\bibitem[Bottinelli et al. (2004a)]{bot04a}
Bottinelli, S., Ceccarelli, C. Lefloch, B., Williams, J.P.,
Castets, A., Caux, E., Cazaux, S., Maret, S., Parise, B., \& Tielens,
A. G. G. M. 2004, ApJ, 615, 354


\bibitem[Bottinelli et al.~(2004b)]{bot04b}
Bottinelli, S., Ceccarelli, C., Neri, R., Williams, J.P., Caux, E.,
Cazaux, S. Lefeloch, B., Maret, S., \& Tielens, A.G.G.M. 2004, ApJ, 617, L69

\bibitem[Caselli et al.~(1999)]{cas99}
Caselli, P., Walmsley, M., Tafalla, M., Dore, L., \& Myers, P. 1999,
ApJ, 523, L165


\bibitem[Cauzax et al.~(2003)]{cau03}
Cazaux, E., Tielens, A.G.G.M., Ceccarelli, C., Castets, A., Wakelam, V.,
Caux, E., Parise, B., \& Teyssier, D., 2003, ApJ, 593, L51

\bibitem[Ceccarelli et al.~(2007)] {cec07}
Ceccarelli, C., Caselli, P., Herbst, E.,
Tielens, A.G.G.M.,  \& Caux, E. 2007 in Protostars and Planets., ed. B. Reipurth, D. Jewitt, K. Keil
(Tucson: University of Arizona Press), 47

\bibitem[Ceccarelli et al.~(2000a)]{cec00a}
Ceccarelli, C., Castets, A., Caux, E., Hollenbach, D., Loinard, L.,
Molinari, S., \& Tielens, A. G. G. M. 2000a, A\&A, 355, 1129

\bibitem[Ceccarelli et al.~(2000b)]{cec00b}
Ceccarelli, C., Loinard, L., Castets, A., Tielens, A.G.G.M., \& Caux, E., 2000b,
A\&A 357, L9

\bibitem[Chandler et al.~(2005)]{cha05} Chandler, C.J., Brogen, C.L.,
Shirley, Y.L. \& Loinard, L. 2005, ApJ, 632, 371

\bibitem[Chiar et al.~(1996)]{chi96} Chiar, J.E., Adamson, A. J., \& Whittet, D.C.B. 1996, ApJ, 472, 665

\bibitem[Collings et al.~(2003)]{cdf03}
Collings, M.P., Dever, J.W., Fraser, H.J., McCoustra, M.R.S., \& Williams, D.A.
2003, ApJ, 583, 1058

\bibitem[Di Francesco et al.~(2007)]{fra07} Di Francesco, J., Evans, N. J. II,
Caselli, P., Myers, P. C., Shirley, Y., Aikawa, Y., Tafalla, M. 2007,
in Protostars and Planets V.,ed. B. Reipurth, D. Jewitt, K. Keil, (Tucson: University of Arizona Press),
17


\bibitem[Doty et al.~(2004)]{dot04} Doty, S. D., Sch\"{o}ier, F. L.,
\& van Dishoeck, E. F. 2004, A\&A, 418, 1021

\bibitem[Ehrenfreund \& Shutte (2000)]{es00}
Ehrenfreund, P., \& Shutte, W. A. 2000, in Astrochemistry: From Molecular
Clouds to Planetary Systems, ed. Y. C. Minh, Y.C. \& E. F. van Dishoeck
(Chelsea, MI; Sheridan Books; Astronomical Society of the Pacific), 135

\bibitem[Fuchs et al.~(2007)]{fuchs07} Fuchs, G. W., Ioppolo, S., Bisschop, S. E., van Dishoeck, E. F., \& Linnartz, H. 2007, A\&A, submitted

\bibitem[Garrod \& Herbst (2006)]{gh06} Garrod, R. T., \& Herbst, E. 2006,
A\&A 457, 927

\bibitem[Garrod et al.~(2006)]{gpc06} Garrod, R. T., Park I-H., Caselli P.,
\& Herbst, E. 2006, Faraday Discussions, 133, 51

\bibitem[Garrod et al.~(2007)]{gwh06} Garrod, R. T., Wakelam, V., \& Herbst, E.
2007 A\&A, 467, 1103


\bibitem[Geppert et al.~(2006)]{gep06} Geppert, W. D., Thomas, R. D.,
Ehlerding, A. et al. 2006, Faraday Discuss. 133, 177

\bibitem[Gibb et al.~(2004)]{gibb04} Gibb. E. L., Rettig, T., Brittain, S., Haywood, R., Simon, T., \& Kulesa, C. 2004, ApJ, 610, L113

\bibitem[Goldsmith \& Arquilla (1985)]{ga85} Goldsmith, P.F. \& Arquilla, R. 1985,
in Protostars and Planets II., ed. D.C. Black, M.S. Matthews
(Tucson: University of Arizona Press), 137

\bibitem[Hasegawa et al.~(1992)]{hhl92} Hasegawa, T.I., Herbst, E. ,\&
Leung, C. M. 1992, ApJS, 82, 167

\bibitem[Hasegawa \& Herbst (1993)]{hh93}
Hasegawa, T. I. \& Herbst, E. MNRAS, 261, 83

\bibitem[Horn et al.~(2004)]{hor04} Horn, A., M$\o$llendal, H., Sekiguchi, O.
et al. 2004, ApJ, 611, 605

\bibitem[Kroes \& Andersson ~(2006)]{ka06}
Kroes, G. J. \& Andersson, S. 2006, in Astrochemistry: Recent Successes and Current Challenges, ed. D. C. Lis, G. A. Blake, E. Herbst (New York: Cambridge Univ. Press) 427



\bibitem[Kuan et al.~(2004)]{kua04} Kuan, Y.-J., Juang, H.-C., Charnley, S.B.,
Hirano, N., Takakuwa, S., Wilner, D.J., Liu, S.-Y., Ohashi, N., Bourke, T.L.,
Qi, C., \& Zhang, Q. 2004, ApJ, 616, L27


\bibitem[Larson (1969)]{lar69}
Larson, R.B. 1969, MNRAS 145, 271

\bibitem[Larson et al.~(2007)]{lar07}
Larsson, B. et al. 2007, A\&A, 466, 999

\bibitem[Lee et al.~(1996)]{lee96} Lee, H-.H., Herbst, E., Pineau des
For\^{e}ts, G., Roueff, E., \& Le Bourlot, J. 1996, A\&A, 311, 690

\bibitem[Lee et al.~(2004)]{lbe04} Lee, J.-E., Bergin, E. A., \&
Evans, N. J. II 2004, ApJ, 617, 360

\bibitem[Lee et al.~(2005)]{leb05} Lee, J.-E., Evans, N. J. II , \&
Bergin, E. A. 2005, ApJ, 631, 351

\bibitem[Loren~(1989)]{lor89}
Loren, R.B. 1989, ApJ, 338, 902

\bibitem[Maret et al.~(2004)]{mar04} Maret, S., Ceccarelli, C., Caux, E.,
Tielens, A.G.G.M., J$\o$rgensen, J.K., van Dishoeck, E.F., Bacmann, A.,
Castets, A., Lefloch, B., Loinard, L., Parise, B. \& Sch\"{o}ier, F. L.
2004, A\&A, 416, 577

\bibitem[Maret et al.~(2005)]{mar05} Maret, S., Ceccarelli, C.,
Tielens, A.G.G.M., Caux, E., Lefloch, B., Faure, A., Castet, A.,
\& Flower, D.R. 2005, A\&A, 442, 527

\bibitem[Masunaga \& Inutsuka (2000)]{mi00} Masunaga, H., \& Inutsuka, S. 2000,
ApJ, 531, 350

\bibitem[Masunaga et al. (1998)]{mi98} Masunaga, H., Miyama, S. M., \&
Inutsuka, S. 1998, ApJ. 495, 346

\bibitem[Millar \& Hatchell (1998)]{mh98} Millar, T. J., \& Hatchell J. 1998,
Faraday Discuss. 109, 15

\bibitem[Omukai (2007)]{omu07} Omukai, K. 2007, PASJ in press

\bibitem[Pontoppidan et al.~(2003)]{ppp03}
Pontoppidan, K. M., Dartois, E., van Dishoeck, E. F., Thi, W. -F., \&
d'Hendecourt, L. 2003, A\&A, 404, L17


\bibitem[Remijian \& Hollis (2006)]{rem06}
Remijian, A.J., \& Hollis, J. M. 2006, ApJ, 640, 842

\bibitem[Ruffle \& Herbst (2000)]{rh00}
Ruffle, D. P., \& Herbst, E. 2000, MNRAS, 319, 837

\bibitem[Ruffle \& Herbst (2001)]{rh01}
Ruffle, D. P., \& Herbst, E. 2001, MNRAS, 324, 1054

\bibitem[Sakai et al. (2007b)]{sak07}
Sakai, N., Sakai, T., \& Yamamoto, S. 2007b, ApSS, in press

\bibitem[Sakai et al. (2007a)]{sshk07}
Sakai, N., Sakai, T., Hirota, T., \& Yamamoto, S. 2007a, ApJ, in press

\bibitem[Sandford \& Allamandola (1988)] {sa88}
Sandford, S. A.,  \& Allamandola, L. J. 1988, Icarus, 76, 201

\bibitem[Schoier et al.~(2002b)]{sch02}
Schoier, F. L., J{\o}rgensen, J. K., van Dishoeck, E. F.,\&  Blake, G. A.
A\&A, 390, 1001

\bibitem[Sch\"{o}ier et al.~(2002a)]{sjv02} Sch\"{o}ier, F. L., J$\o$rgensen,
J. K., van Dishoeck, E. F.,  \& Blake, G. A. 2002, A\&A 391, 1001

\bibitem[Shu (1977)]{shu77} Shu, F. H. 1977, ApJ, 214, 488

\bibitem[Tachihara et al.~(2000)]{tac00}
Tachihara, K., Mizuno, A., \& Fukui, Y. 2000, ApJ, 528, 817

\bibitem[Tafalla et al.~(2002)]{taf02}
Tafalla, M., Myers, P. C., Caselli, P., Walmsley, C. M., \& Comito,
C. 2002, ApJ, 569, 815

\bibitem[Vastel et al.~(2004)]{vas04}
Vastel, C., Phillips, T.G., Yoshida, H. 2004, ApJ, 606, 127

\bibitem[Viti \& Williams (1999)]{vw99} Viti, S., \& Williams, D. A. 1999,
MNRAS, 305, 755

\bibitem[Viti et al.~(2004)]{vit04} Viti, S., Collings, M. P., Dever, J. W.,
McCoustra, M. R. S,. \& Williams, D. A. 2004, MNRAS, 354, 1141

\bibitem[Wakelam et al.~(2004)]{wakelam04}
Wakelam, V., Caselli, P., Ceccarelli, C., Herbst, E., \& Castets, A. 2004, A\&A, 422, 159

\bibitem[Walker et al.~(1986)]{wal86}
Walker, C. K., Lada, C.J., Young, E.T., Maloney, P.R., \& Wilking, B.A. 1986,
ApJ, 309, L47

\bibitem[Watanabe \& Kouchi (2002a)]{wkl02} Watanabe, N.,  \& Kouchi, A. 2002a,
ApJ, 571, L173, 

\bibitem[Watanabe \& Kouchi (2002b)]{wk02}
Watanabe, N, \& Kouchi, A. 2002b, ApJ, 567, 651

\bibitem[Watanabe et al.~(2007)]{wm07}
Watanabe, N., Mouri, O., Nagaoka, A., Kouchi, A., \& Pirronello, V. 2007,
ApJ, in press



\bibitem[Williams (1968)]{wil68} Williams, D.A. 1968, ApJ, 151, 935





\end{thebibliography}
\end{document}